\documentstyle[preprint,revtex]{aps}
\begin{document}
\draft
\preprint{UCB-PTH-92/23}
\preprint{LBL-32498}
\vskip -45pt
June, 1992
\vskip 35pt
\begin{title}
Radiative corrections to the Higgs boson decay rate \\
$\Gamma(H\rightarrow ZZ)$ in the minimal supersymmetric model
\end{title}
\author{Damien Pierce and Aris Papadopoulos}
\begin{instit}
Department of Physics, University of California, Berkeley\\
and\\
Lawrence Berkeley Laboratory, 1 Cyclotron Road, Berkeley, CA 94720.
\end{instit}
\pagestyle{empty}
\begin{abstract}
We consider radiative corrections to the decay rate
$\Gamma(H\rightarrow ZZ)$ of the heavy {\it CP}-even Higgs
boson of the minimal supersymmetric model to two $Z$ bosons.
We perform a one loop Feynman diagram calculation
in the on-mass-shell renormalization scheme, and include the third
generation of quarks and squarks. The tree level rate
is suppressed by a mixing angle factor and decreases as $1/M_H$
for large $M_H$. The corrected rate overcomes this suppression and
increases with $M_H$ for $M_H
\raise.3ex\hbox{$>$\kern-.75em\lower1ex\hbox{$\sim$}} 500$~GeV.
The corrections can be very large and depend in detail on
the top squark masses and $A$-term, as well as the supersymmetric
Higgs mass parameter $\mu$.
\end{abstract}
Typeset in {\it REVTEX}
\vfill
\vskip .35in
\vbox{\footnotesize\noindent This work was supported by the
Director, Office of Energy Research, Office of
High Energy and Nuclear Physics, Division of High Energy Physics
of the U.S. Department of Energy under contract DE-AC03-76SF00098 and in part
by the National Science Foundation under grant PHY--90--21139.}
\pagebreak
\section{Introduction}
One of the least attractive features of the standard model
(SM) is the existence of the naturalness problem. Roughly speaking
this means that when one
computes corrections to the Higgs boson mass one finds quadratically
divergent
contributions. This situation implies that input parameters must be
extremely
fine-tuned at high energies to yield the low energy physics that we
observe, a situation that is unappealing especially in connection
with GUTs.

\setcounter{page}{1}
\pagestyle{plain}
One way to control the naturalness problem is to consider
supersymmetric
(SUSY) extensions of the standard model. Here the quadratic
divergences are
cancelled by loop diagrams involving the superpartners of the SM
particles.
We know that SUSY must be broken in the real world, and yet the
scale of
supersymmetry breaking must not be too large or the hierarchy
problem will be
reintroduced. Thus, although superparticles must be sufficiently
heavy to have avoided detection at present colliders, they cannot be
much
heavier than a few TeV if we are to meet the naturalness criterion.

In this work we will be concerned with the simplest supersymmetric
extension of the standard model (MSSM)
\cite{Haber Kane,MSSM papers}.
In the MSSM we need two Higgs doublets $H_1$ and $H_2$ to give
masses
to up and down type fermions and to assure cancellation of
anomalies. The
neutral Higgs spectrum consists of two {\it CP}-even Higgs scalar
particles $H$ and $h$ (where $M_H>M_h$),
one {\it CP}-odd particle $A$, and a Goldstone boson $G$ which is
``eaten by'' and gives mass to the $Z$ boson. The Higgs sector
of the MSSM is highly constrained. At tree level the
Higgs boson masses and couplings are determined by two input
parameters.
We take these to be the mass of the {\it CP}-odd Higgs boson $M_A$
and an angle $\beta$ which at tree level is given by
$\tan\beta=v_2/v_1$
where $v_2$ and $v_1$ are the vacuum expectation values of the
two Higgs boson fields $H_2$ and
$H_1$. The tree level masses of the {\it CP}-even Higgs bosons are
then given by
\begin{equation}
M_{H,h}^2= {1\over 2}\biggl(M_A^2+M_Z^2
\pm\sqrt{(M_A^2+M_Z^2)^2-4M_Z^2 M_A^2\cos^2(2\beta)}\biggr).
\eqnum{1}
\end{equation}
The above equation implies the inequalities $M_h<M_Z,\;\; M_H>M_Z$
and the sum rule $M_H^2+M_h^2=M_Z^2+M_A^2$.

Recently it was shown that one loop corrections involving top-quark
and squark loops can
significantly modify the sum rule \cite{Berger} and also violate
the bound $M_h<M_Z$ \cite{Pierce,Higgs papers}. For
1~TeV squark masses the correction to the light Higgs boson mass is
of the order 20 (50)~GeV for a top mass of 150 (200)~GeV.
Corrections to the neutral Higgs boson mass sum rule
due to the gauge-Higgs and gaugino-higgsino sectors were considered
earlier \cite{Gunion and Turski} and were found to be generically
small.

In this work we consider corrections to the decay rate
$\Gamma(H\rightarrow ZZ)$ which is relevant for the detection of
the heavy
Higgs boson at a proton supercollider such as the SSC via the
``gold-plated'' mode
$H\rightarrow ZZ\rightarrow \ell^+\ell^-\ell^+\ell^-$,
where $\ell$ is $e$ or $\mu$. We
confine ourselves to corrections due to third family
(top and bottom) quark and squark loops.
Previous work on this subject has appeared in Ref.'s \cite{Barger} and
\cite{Gunion et al} where the effective potential and the
renormalization group methods are
used. We perform a Feynman diagram calculation utilizing the
on-mass-shell
renormalization scheme, and present explicit analytic results.
The structure of the paper is as follows: in Section 2 we present
our renormalization procedure, in Section 3 we discuss our results,
Section 4 lists briefly our conclusions, and in
the Appendix we present the necessary explicit formulas.

\section{Formalism for radiative corrections}
Due to the presence of mixing in the {\it CP}-even and
{\it CP}-odd sectors
the renormalization of the Higgs sector of the MSSM presents a few
complications when compared to the standard model. Therefore, in this
section we present in detail our renormalization procedure. We
follow
the approach of Aoki {\it et. al.}\ \cite{Aoki} adapted to the MSSM.

The Higgs potential in the MSSM is
\begin{eqnarray}
V={g^2+g'^2\over8} && \left(H^{i*}_1H^i_1-H^{i*}_2H^i_2\right)^2
 + {g'^2\over2}|H_1^{*i}H_2^i|^2
\qquad\qquad\eqnum{2}\\
&&+(m^2_1+\mu^2)H^{i*}_1H^{i}_1+(m^2_2+
\mu^2)H^{i*}_2H^{i}_2
-\left(m^2_3\epsilon_{ij}H^{i}_1H^{j}_2+h.c.\right),\nonumber
\end{eqnarray}
where $g$($g'$) is the $SU(2)_L(U(1)_Y)$ gauge coupling, the
$m_{{\rm i}}$'s, (i = 1,2,3) are the soft supersymmetry breaking Higgs
sector mass parameters, and $\mu$ is the supersymmetric Higgs mass
parameter.
We can absorb $\mu^2$ in Eq.(1) by redefining
$m_1^2+\mu^2\rightarrow m_1^2$
and similarly for $m_2^2.$
$H_1$ and $H_2$ are given in terms of the shifted (but unrotated)
fields by
$$
H_1= {1\over \sqrt{2}} \left(\matrix{v_1+S_1-iP_1\cr
               \sqrt{2}H_{-}\cr}\right),\qquad\qquad
H_2={1\over \sqrt{2}} \left(\matrix{\sqrt{2}H_{+}\cr
                    v_2+S_2+iP_2\cr}\right).
$$
In order to discuss the tadpole and mixing structure of the theory we
need the terms that are linear and quadratic in $S_1$, $S_2$ and
quadratic in $P_1$, $P_2$. These are given by
\begin{eqnarray}&&V_s=\left({g^2+g'^2\over8}
(v_1^2-v_2^2)v_1+m_1^2v_1-m_{3}^2v_2\right)S_1
+ \left({g^2+g'^2\over 8}(v_2^2-v_1^2)v_2+m_2^2v_2-m_{3}^2v_1\right)S_2
\nonumber\\
&&\qquad+\;\left({g^2+g'^2\over 16}(3v_1^2-v_2^2)+{m_1^2\over 2} \right)S_1^2
+\left({g^2+g'^2\over 16}(3v_2^2-v_1^2)+{m_2^2\over 2}\right)S_2^2\nonumber\\
&&\qquad-\;\left({g^2+g'^2\over 4}v_1v_2+m_3^2\right)S_1S_2\eqnum{3a}\\
&&V_p=\left({g^2+g'^2\over 16}(v_1^2-v_2^2)+{m_1^2\over 2}\right)P_1^2
+\left({g^2+g'^2 \over 16}(v_2^2-v_1^2)+{m_2^2\over 2}\right)P_2^2
-m_{3}^2P_1P_2.\eqnum{3b}
\end{eqnarray}
We now define the coefficients of $S_1$ and $S_2$ in Eq.(3a) to be
\begin{eqnarray}T_1
                  &={g^2+g'^2 \over 8}(v_1^2-v_2^2)v_1
                     +m_1^2v_1-m_{3}^2v_2\eqnum{4a} \\
             T_2
                  &={g^2+g'^2 \over 8}(v_2^2-v_1^2)v_2
                     +m_2^2v_2-m_{3}^2v_1.\eqnum{4b}
\end{eqnarray}
Eliminating $m_1^2$, $m_2^2$ in favor of $T_1$, $T_2$ from
Eqs.(4) and substituting back in Eqs.(3) we obtain,
using a matrix notation
\begin{eqnarray}
V_{s}
=(S_1&&\:S_2)\left(\matrix{T_1\cr T_2\cr}\right)
+{1\over 2}(S_1\:S_2)\left(\matrix{{T_1\over v_1}&0\cr
0&{T_2\over v_2}\cr}\right)\left( \matrix{S_1\cr S_2\cr}\right)
\eqnum{5a}\\
&&+{1 \over 2}(S_1\:S_2)\left(
\matrix{{g^2+g'^2\over 4}v_1^2+m_{3}^2{v_2\over v_1}
&-{g^2+g'^2\over4}v_1v_2-m_{3}^2\cr-{g^2+g'^2\over 4}v_1v_2-m_{3}^2
& {g^2+g'^2\over4}v_2^2+m_{3}^2{v_1\over v_2}\cr  }\right)
\left(\matrix{S_1\cr S_2\cr}\right)\nonumber
\end{eqnarray}
\begin{equation}
V_{p}={1\over 2}(P_1\:P_2)\left(\matrix{{T_1\over v_1}&0\cr
                              0&{T_2\over v_2}\cr} \right)
\left(\matrix{P_1\cr P_2\cr}\right)
+{1 \over 2}(P_1\:P_2)\left(\matrix{m_3^2
{v_2\over v_1}&-m_{3}^2\cr
-m_3^2&m_3^2{v_1\over v_2}\cr}\right)
\left(\matrix{P_1\cr P_2\cr}\right).\eqnum{5b}
\end{equation}
\noindent The next step is to introduce rotation matrices
$O(\alpha)$ and
$O(\beta)$ such that the part of the {\it CP}-even and {\it CP}-odd
mass matrices that does not depend on $T_1$, $T_2$ is
diagonalized. Specifically, by defining
$$\left(\matrix{S_1\cr
                S_2\cr}\right)=O(\alpha)
\left(\matrix{H\cr h\cr}\right)=
 \left(\matrix{\cos\alpha&-\sin\alpha\cr
               \sin\alpha&\cos\alpha\cr}\right)
\left(\matrix{H\cr
              h\cr}\right)$$
and
$$\left(\matrix{P_1\cr P_2\cr}\right)=
O(\beta)\left(\matrix{G\cr A\cr}\right)=
\left(\matrix{\cos\beta&-\sin\beta\cr\sin\beta&\cos\beta\cr}\right)
\left(\matrix{G\cr A\cr}\right)$$
we find that
\begin{equation}
V_s=(H\:\:h)\left(\matrix{T_H\cr
                          T_h\cr}\right)
+{1\over 2}(H\:\:h)O(-\alpha)\left(
\matrix{{T_1\over v_1}&0\cr 0&{T_2\over v_2}\cr}
\right)
O(\alpha)\left(
\matrix{H\cr h\cr}
\right)
+{1\over 2}(H\:h)\left(   \matrix{M_H^2&0\cr 0&M_h^2\cr}
  \right)    \left(   \matrix{H\cr h\cr}   \right)\eqnum{6a}
\end{equation}
\begin{equation}
V_p={1\over 2}(G\:A)O(-\beta)
\left(\matrix{{T_1\over v_1}&0\cr
      0&{T_2\over v_2}\cr}\right)
O(\beta)\left(\matrix{G\cr A\cr}\right)
+{1\over2}(G\:A)\left(\matrix{0&0\cr 0&M_A^2\cr}\right)
\left(\matrix{G\cr A\cr}\right).\eqnum{6b}
\end{equation}
Here we have defined
$$\left(\matrix{T_1\cr T_2\cr}\right)=
O(\alpha)\left(\matrix{T_H\cr
T_h\cr}\right).$$
The parameters $\beta$, $\alpha$, $M_H$, $M_h$ and $M_A$
are related to the original fundamental parameters $v_1$, $v_2$ and
$m_3^2$ by the following formulas
\begin{equation}
\tan\beta={v_2 \over v_1},\qquad
M_A^2 = m_3^2\left(\tan\beta + \cot \beta\right),\qquad
\tan{2\alpha}={M_A^2+M_Z^2 \over M_A^2-M_Z^2}\tan2\beta,
\eqnum{7}
\end{equation}
as well as Eq.(1).
Here we used $M_Z^2={g^2+g'^2\over4}(v_1^2+v_2^2)$.
Carrying out the remaining matrix multiplications involving the
tadpole contributions to the mass matrices
 we obtain the final result
\begin{equation}
V_s=HT_H+hT_h
+{1\over 2}(H\:h)
\left(\matrix{M_H^2+b_{HH}&b_{Hh}\cr
 b_{Hh}&M_h^2+b_{hh}\cr}\right)
\left(\matrix{H\cr h\cr}\right)\eqnum{8a}
\end{equation}
\begin{equation}
V_p={1\over 2}(G\:A)\left(\matrix{b_{GG}&b_{GA}\cr
b_{GA}&M_A^2+b_{AA}\cr}\right)
\left(\matrix{G\cr A\cr}\right)\eqnum{8b}
\end{equation}
with
$$
b_{HH}={2\over v\sin 2\beta}
\left(T_H(\cos^{3}\alpha\sin\beta+
\sin ^{3}\alpha\cos\beta)\right.
\left.+T_h\sin\alpha\cos\alpha\sin(\alpha-\beta)\right)
$$
\begin{equation}
b_{Hh}={\sin 2\alpha\over v\sin 2\beta}
\left(T_H\sin(\alpha-\beta)
+T_h\cos(\alpha-\beta)\right)\eqnum{8c}
\end{equation}
$$
b_{hh}={2\over v\sin 2\beta}
\left(T_H\cos\alpha\sin\alpha\cos(\alpha-\beta)\right.
\left.+T_h(\cos^{3}\alpha\cos\beta
-\sin ^{3}\alpha\sin\beta)\right)
$$
and
$$b_{GG}={1\over v}\left(T_H\cos(\alpha-\beta)
-T_h\sin(\alpha-\beta)\right)$$
\begin{equation}
b_{GA}={1\over v}\left(T_H\sin(\alpha-\beta)
+T_h\cos(\alpha-\beta)\right)
\eqnum{8d}
\end{equation}
$$b_{AA}={2\over v\sin 2\beta}
\left(T_H(\sin^3\beta\cos\alpha
+\cos^3\beta\sin\alpha)+T_h(\cos^3\beta\cos\alpha
-\sin^3\beta\sin\alpha)\right)$$
The terms linear in $H$ and $h$ are to be thought
 of as counterterms for the
tadpoles. To each order in the loop expansion we require that the
total tadpole contribution vanishes. At tree level this implies
$-iT_H=0=-iT_h$. This then
gives the conventional tree level masses. At one loop $-iT_H$
($-iT_h$) must cancel the one loop $H$ ($h$) tadpole diagrams
$i\tau_H$ ($i\tau_h$) (Fig.1).
These conditions determine $T_H$ and $T_h$ and Eqs.(8)
determine their contribution to the one loop mass matrices.

Taking as renormalized inputs $\tan\beta$
and $M_A$ we calculate
the physical masses $M_H,\; M_h$ and the decay rate
$\Gamma (H\rightarrow ZZ)$
at one loop. It follows that the measurement of any two of the
physical quantities $M_A,\; M_H, \; M_h$ and
$\Gamma (H\rightarrow ZZ)$ will
allow us to make a prediction for the other two. We stress that
$\beta$ is only to be viewed as a useful parametrization of physical
observables. Since by itself $\beta$ has no physical meaning we can
renormalize it in any suitably convenient way. We explain our
renormalization prescription for $\beta$ below.

{}From this point on we adopt the following notation conventions:
a quantity such as a field, coupling, or mass with a subscript $0$
indicates a bare quantity, renormalized quantities have a subscript
$r$, and
physical observables such as the pole of a propagator do not have
subscripts. The bare tree Lagrangian contains
\begin{eqnarray}
{\cal L}\supset {1\over 2}\partial_{\mu} &&H_0\partial^{\mu} H_0
+{1\over 2}\partial_{\mu} h_0\partial^{\mu} h_0\qquad\eqnum{9}\\
&& -{1\over 2}(M_{H_0}^2+b_{HH})H_0^2
-{1\over 2}(M_{h_0}^2+b_{hh})h_0^2-b_{Hh}H_0h_0\nonumber
\end{eqnarray}
where $M_{H_0}^2$ and $M_{h_0}^2$
are taken to be functions of  $M_{A_0},\; \beta_0$ and $M_{Z_0}$ as
given by
equation (1). We now write the bare parameters in terms of
renormalized parameters and shifts
\begin{equation}
\beta_0=\beta_r+\delta\beta,\;\;\;M_{A_0}^2=
M_{A_r}^2+\delta M_A^2,\;\;\;M_{Z_0}^2=M_{Z_r}^2
+\delta M_Z^2\eqnum{10}
\end{equation}
and also introduce wave function renormalization
\begin{equation}
H_0=Z_{HH}^{1\over2} H_r+Z_{Hh}^{1\over2} h_r,\qquad
h_0=Z_{hh}^{1\over2} h_r+Z_{hH}^{1\over2} H_r.\eqnum{11}
\end{equation}
Note that $Z_{HH}^{1\over2} =1+{\cal O}(\alpha),\;
Z_{hh}^{1\over2}=1+{\cal O}(\alpha)$ while $Z_{Hh}^{1\over2},
\;Z_{hH}^{1\over2},\;b_{HH},\;b_{Hh},$ and $b_{hh}$
are all ${\cal O}(\alpha)$.
Substituting equations (10) and (11) into (9) we obtain
the one loop renormalized two-point functions
$$i\Gamma_{HH}(p^2)=(Z_{HH}^{1\over2})^2(p^2-M_{H_r}^2)
-\frac{\partial M_{H_r}^2}{\partial x_{ir}}\delta x_{i}
-b_{HH}+\Pi_{HH}(p^2)$$
\begin{equation}
i\Gamma_{hh}(p^2)=
(Z_{hh}^{1\over2})^2(p^2-M_{h_r}^2)
-\frac{\partial M_{h_r}^2}{\partial x_{ir}}
\delta x_i-b_{hh}+\Pi_{hh}(p^2)\eqnum{12}
\end{equation}
$$i\Gamma_{Hh}(p^2)=Z_{Hh}^{1\over2}(p^2-M_{H_r}^2)
+Z_{hH}^{1\over2}(p^2-M_{h_r}^2)-b_{Hh}+\Pi_{Hh}(p^2),$$
where $x_{ir}=M_{A_r}^2,\; M_{Z_r}^2,\; \beta_r$ and
the $\Pi$'s are the scalar self-energies (Fig.2).
The on-shell renormalization conditions are \cite{Aoki}
$$i\Gamma_{HH}(M_H^2)=i\Gamma_{Hh}(M_h^2)=
  i\Gamma_{Hh}(M_H^2)=i\Gamma_{hh}(M_h^2)=0$$
\begin{equation}
i\left.{\partial\Gamma_{HH}\over \partial p^2}\right|_{p^2=M_H^2}=1=
i\left.{\partial\Gamma_{hh}\over \partial p^2}\right|_{p^2=M_h^2}
\eqnum{13}
\end{equation}
Here $M_H$ and $M_h$ are the physical masses of $H$ and $h$.
Making the definitions $\delta M_H^2=\Pi_{HH}(M_{H_r}^2)-b_{HH}$
and similarly
for $\delta M_h^2$, we obtain from Eqs.(12) and (13)
\begin{equation}
M_H^2=M_{H_r}^2+\frac{\partial M_{H_r}^2}{\partial x_{ir}}
\delta x_{i}-\delta M_H^2\eqnum{14a}
\end{equation}
\begin{equation}
M_h^2=M_{h_r}^2+\frac{\partial M_{h_r}^2}{\partial x_{ir}}
\delta x_{i}-\delta M_h^2\eqnum{14b}
\end{equation}
\begin{equation} Z_{HH}^{1\over2} =1-{1\over 2}\Pi_{HH}'(M_{H_r}^2)
\eqnum{14c}
\end{equation}
\begin{equation}
Z_{hh}^{1\over2} =1-{1\over 2}\Pi_{hh}'(M_{h_r}^2) \eqnum{14d}
\end{equation}
\begin{equation} Z_{hH}^{1\over2}={1\over M_{H_r}^2- M_{h_r}^2}\left(
-\Pi_{Hh}( M_{H_r}^2)+b_{Hh}\right)\eqnum{14e}
\end{equation}
\begin{equation}
Z_{Hh}^{1\over2}={1\over M_{h_r}^2- M_{H_r}^2}\left(
-\Pi_{Hh}( M_{h_r}^2)+b_{Hh}\right)\eqnum{14f},\end{equation}
where the prime in Eqs.(14c,d) indicates differentiation with
respect to $p^2$.
Note that $ M_{H_r}^2$ and $ M_{h_r}^2$ have the same functional form
as in Eq.(1) except that they are functions of renormalized
quantities, {\it i.e.}
\begin{equation}
M_{{H_r},{h_r}}^2={1\over 2}\biggl(M_{A_r}^2+M_{Z_r}^2
\pm\sqrt{(M_{A_r}^2+M_{Z_r}^2)^2
-4M_{A_r}^2M_{Z_r}^2\cos^2(2\beta_r)}
\biggr).\eqnum{15}
\end{equation}
We now drop the subscript $r$
on $M_{Z_r},\; M_{A_r}$ and $\beta_r$.
Eqs.(14a,b) determine the physical {\it CP}-even Higgs
 boson masses in
terms of self energies, tadpole contributions, and
shifts of the inputs parameters $\delta x_{i}$.
 We now determine the
shifts. The shift $\delta M_A^2$ is defined so that
 $M_A$ is equal to the
physical $A$ mass. An analysis similar to that of the
 {\it CP}-even sector yields
\begin{equation}
\delta M_A^2=\Pi_{AA}(M_A^2)-b_{AA}.\eqnum{16}
\end{equation}
Additionally, we find for the shift in the $Z$-boson mass
\begin{equation}
 \delta M_Z^2=\Pi_{ZZ}^T(M_Z^2)\eqnum{17}
\end{equation}
where $\Pi_{ZZ}^T$ is the transverse part of the $Z$ boson
self energy,
$\Pi_{ZZ}^{\mu\nu}=g^{\mu\nu}\Pi_{ZZ}^T+
{p^{\mu}p^{\nu}\over p^2}\Pi_{ZZ}^{L}$.
At this point it is worth noting that if we are only
 interested in the sum
$M_H^2+M_h^2$ we do not need a specification for $\delta\beta\;$.
When Eqs.(14a)
and (14b) are added the terms proportional to
$\delta \beta$ cancel leaving
\begin{eqnarray}
M_H^2+M_h^2=M_A^2+M_Z^2&&
-\Pi_{HH}(M_H^2)-\Pi_{hh}(M_h^2)+
\Pi_{AA}(M_A^2)+\Pi_{ZZ}^T(M_Z^2)\nonumber\\
&&+b_{HH}+b_{hh}-b_{AA}\eqnum{18}
\end{eqnarray}
This is just the renormalization of the neutral Higgs boson mass
sum rule and the divergences in Eq.(18)
implicit in the $\Pi$'s and $b$'s cancel leaving behind a finite
correction. Since we demand that $M_H$ and $M_h$ are physical masses
they must be individually finite. Equivalently, since
$M_H^2+M_h^2$ is finite we must have that $M_H^2-M_h^2$
is also free of divergences. This latter requirement
gives
\begin{equation}
\frac{\partial \Delta}{\partial \beta}
\delta\beta+\frac{\partial \Delta}{\partial M_Z^2}
\delta M_Z^2+\frac{\partial \Delta}{\partial M_A^2}\delta M_A^2
-\delta M_H^2+\delta M_h^2=\mbox{finite}\eqnum{19}
\end{equation}
where $\Delta=\sqrt{(M_A^2+M_Z^2)^2-4M_A^2M_Z^2\cos^2(2\beta)}.$
The above equation clearly determines only the ``infinite'' part of
$\delta\beta$. By ``infinite'' we mean the part
that is proportional to
$C_{UV}={1\over \epsilon}-\gamma +\log4\pi\;$
in dimensional regularization.
To fully specify $\delta\beta$ we take a
$\overline{MS}$-type approach and
define $\delta\beta$ to be purely ``infinite'' so that
Eq.(19) becomes
\begin{eqnarray}
\frac{\partial \Delta}{\partial \beta}\delta\beta=
-\left(\frac{\partial \Delta}{\partial M_Z^2}
\delta M_Z^2+\frac{\partial \Delta}{\partial M_A^2}
\delta M_A^2-\delta M_H^2+\delta M_h^2\right)_{\infty}\eqnum{20}
\end{eqnarray}
where the subscript $\infty$ on a quantity indicates the ``infinite''
part of that quantity. Eq.(20) implies
\begin{eqnarray}
\delta\beta=&&{1\over 2M_A^2M_Z^2\sin(4\beta)}\eqnum{21}\\
&&\times\biggl(\left(M_Z^2\;\delta M_A^2+M_A^2\;\delta M_Z^2\right)
\cos^2(2\beta)-\,M_H^2\;\delta M_h^2\,-\,M_h^2\;\delta M_H^2\biggr)_{\infty}.
\nonumber
\end{eqnarray}
This definition of $\beta$ at one loop gives renormalized
{\it CP}-even
Higgs boson masses in close agreement with those obtained using
the effective potential \cite{Pierce}.
This shift in $\delta\beta$ induces a shift in
$\alpha$ through equation (7)
\begin{equation}
\delta\alpha=\sin(4\alpha)\left({\delta\beta\over \sin(4\beta)}+
{M_A^2\;\delta M_Z^2-M_Z^2\;\delta M_A^2\over 2(M_A^4-M_Z^4)}
\right).\eqnum{22}
\end{equation}
We now come to the renormalization of the $HZZ$ coupling.
The bare $HZZ$ and $hZZ$ couplings are given by
$$\lambda^{HZZ}_0={e_0M_{Z_0}^{3}\over M_{W_0}
(M_{Z_0}^2-M_{W_0}^2)^{1\over2}}
\cos(\beta_0-\alpha_0),\qquad
\lambda^{hZZ}_0=\lambda^{HZZ}_0\tan(\beta_0-\alpha_0).$$
Defining
$$e_0=e_r+\delta e,\qquad Z_{HH}^{1\over2}
=1+\delta Z_{HH}^{1\over2},\qquad
(Z_{ZZ}^{1\over2})^2=1+\delta Z_{ZZ}$$
(here $Z_{\mu_0}=Z_{ZZ}^{1\over 2}Z_{\mu_r}
+Z_{ZA}^{1\over 2}A_{\mu_r}$
where $Z_{\mu_0}(Z_{\mu_r})$
is the bare (renormalized) $Z$ boson field
and $A_{\mu_r}$ is the renormalized photon field)
we obtain for the renormalized one loop 3-point function
\begin{equation} \Gamma^{HZZ}_{\mu\nu}=
(\lambda^{HZZ}_r+\lambda^{HZZ}_{CT})g_{\mu\nu}
+\Delta\Gamma^{HZZ}_{\mu\nu}\eqnum{23}\end{equation}
where $\lambda^{HZZ}_r={e_rM_{Z_r}^{3}\cos(\beta_r-\alpha_r)\over
M_{Wr}(M_{Z_r}^2-M_{Wr}^2)^{1\over2}}$ and
\begin{eqnarray}
\lambda^{HZZ}_{CT}=\lambda^{HZZ}_r\biggl({\delta e\over e}&&
+{3\over 2}{\delta M_Z^2\over
M_Z^2}-{1\over 2}{\delta M_Z^2-\delta M_W^2\over M_Z^2-M_W^2}
-{1\over 2}{\delta M_{W}^2\over M_{W}^2}\eqnum{24}\\
&&-\tan(\beta_r-\alpha_r)(\delta\beta-\delta\alpha)
+\delta Z_{HH}^{1\over2}+\delta Z_{ZZ}+Z_{hH}^{1\over2}
\tan(\beta_r-\alpha_r)\biggr)\nonumber
\end{eqnarray}
and $\Delta\Gamma^{HZZ}_{\mu\nu}$ is the explicit one
loop Feynman diagram contribution (Fig.3).
The angle $\alpha_r$ is defined as in
Eq.(7), but with the right hand side written in terms of
renormalized quantities.
The expressions for $\delta M_Z^2,\;\delta\beta,\;\delta\alpha,
\;Z_{HH}^{1\over2}$ and $Z_{hH}^{1\over2}$ in terms
of self energies and tadpole contributions
are given in Eqs.(17), (21), (22) and (14c,e). We
simply state the results for the remaining shifts
$\delta e,\;\delta M_W^2$ and
$\delta Z_{ZZ}$. We have
\begin{eqnarray}
&&{\delta e\over e}={1\over2}{\Pi_{\gamma\gamma}^T}'(0)
+ \left({4c^2_{\mbox{$\scriptscriptstyle W$}}
-3\over 4s_{\mbox{$\scriptscriptstyle W$}}
c_{\mbox{$\scriptscriptstyle W$}}}\right)
{\Pi_{Z\gamma}^T(0)\over M_Z^2},\eqnum{25}\\
&&\delta M_{W}^2=\Pi_{WW}^T(M_{W}^2),\qquad
\delta Z_{ZZ}=-{\Pi_{ZZ}^T}'(M_Z^2)\nonumber
\end{eqnarray}
where $c_{\mbox{$\scriptscriptstyle W$}} = M_W/M_Z$ and
$s_{\mbox{$\scriptscriptstyle W$}}=
\sqrt{1-c^2_{\mbox{$\scriptscriptstyle W$}}}.$
We note that $\Pi_{Z\gamma}^T(0)$ vanishes in our case.
The $H\!\!-\!\!h$ mixing gives a contribution to
$\Gamma_{HZZ}^{\mu\nu}$ through the term proportional to
$Z_{hH}^{1\over2}$. The
quantity on the R.H.S. of Eq.(23) is given as a sum of terms
which are individually divergent.
In the full sum the divergences must of course cancel.
We checked both analytically and numerically that
this is indeed the case. The
renormalizability of the theory
requires that the definition of $\delta\beta$ which renders the
{\it CP}-even Higgs boson masses finite also gives finite couplings.

The explicit one loop Feynman diagrams shown in Fig.3
give a contribution to
the three-point function which can be expanded in
terms of form factors as
\begin{equation}
\Delta\Gamma_{HZZ}^{\mu\nu}=D_0g^{\mu\nu}+D_1p_1^{\mu}p_1^{\nu}
+D_2p_2^{\mu}p_2^{\nu}+D_3p_1^{\mu}p_2^{\nu}+
D_4p_2^{\mu}p_1^{\nu}\eqnum{26}
\end{equation}
(a form factor proportional to
$\epsilon^{\mu\nu\alpha\beta}p_{1\alpha}p_{2\beta}$
vanishes by {\it CP}\ \ invariance).
The formula for the decay rate at one loop is
\begin{eqnarray} \Gamma ={\sqrt{1-4r}\over128\pi r^2 M_H}\Biggl\{
&&(1-4r+12r^2)\biggl(
(\lambda^{HZZ}_r)^2\;+\;2\lambda^{HZZ}_r{\rm Re}(\lambda_{CT}^{HZZ}+D_0)+
|\lambda_{CT}^{HZZ}\;+\;D_0|^2\biggr) \nonumber\\
&&+\;\;M_H^2(1-2r)(1-4r)\biggl(\lambda^{HZZ}_r\,{\rm Re}(D_4)+
{\rm Re}\bigl[(\lambda_{CT}^{HZZ}+D_0)D_4^*\bigr]\biggr)\nonumber\\
&&+\;\; M_H^4\left({1\over 2} -2r\right)^2|D_4|^2\Biggr\}\eqnum{27}
\end{eqnarray}
where $r=M_Z^2/M_H^2$ and we list $D_0$ and $D_4$ in the Appendix.

We note  that the terms in the above expression which do not
involve $\lambda_r^{HZZ}$  are formally of ${\cal O}(g^6)$.
Nevertheless
we find that for large Higgs boson mass ($M_H\gg M_Z$)
they are numerically important. This is because
$\lambda_r^{HZZ}$ is proportional to $\cos (\alpha-\beta)$
which is proportional to $1/M_H^2$ for large $M_H$ and hence small.
Keeping these ${\cal O}(g^6)$ terms is
consistent: the terms in the amplitude that are of ${\cal O}(g^5)$
which arise at two loop level also
give a contribution of ${\cal O}(g^6)$ in the decay rate, but these
two loop ${\cal O}(g^6)$ terms are
proportional to $\cos (\alpha-\beta)$ and are thus suppressed when
$M_H\gg M_Z$,
in precisely the region where the ${\cal O}(g^6)$ terms in our one
loop
expression become large.
\section{Results}
In the MSSM at tree level the decay rate $\Gamma(H\rightarrow ZZ)$
is suppressed relative to the
same decay rate in the standard model by the factor
$\cos^2 (\alpha-\beta)$.
The ``gold-plated'' decay mode $H\rightarrow ZZ\rightarrow 4\ell$
has great
discovery potential for a standard model Higgs boson at a proton super
collider such as the SSC for Higgs boson
masses 130~GeV~$\raise.3ex\hbox{$<$\kern-.75em\lower1ex\hbox{$\sim$}}
{}~\;M_{\phi}\;~\raise.3ex\hbox{$<$\kern-.75em\lower1ex\hbox{$\sim$}}$
800~GeV \cite{Higgs Hunter}.
The discovery potential for the heavy Higgs boson of the MSSM in
this mode
is not as promising due to the above mentioned suppression factor.
However, the ``gold-plated'' mode
may be the only discovery mode for the heavy Higgs boson at a hadron
collider \cite{Gunion91}.
The discovery potential is improved
when radiative corrections are taken into account.

We discuss our numerical results below. We have checked our numerics
in a number of ways. First, we checked the cancellation of divergences
as mentioned in the last section. Second, we found our result for
the correction to the neutral Higgs boson mass sum rule agreed very
closely with that of Ref.\cite{Berger}. Lastly, we checked that our
calculation, when modified to give the correction to the standard
model Higgs boson decay rate to two $Z$'s due to an extra heavy
fermion doublet, agrees with the results of Ref.\cite{Kniehl}.

In Fig.4a we show the tree level and radiatively corrected decay rate
versus the heavy Higgs mass for
$\tan\beta$=5 and a top quark mass of 160~GeV. In this
figure we have not included mixing effects,
{\it i.e.} $A_t=A_b=\mu=0$ and
the squark masses are all equal. We show the corrected rate for the
two squark mass choices
$M_{sq}=300$~GeV and $M_{sq}=1000$~GeV. We see in Fig.4a the
importance of
keeping corrections which are of ${\cal O}(g^6)$ in the rate.
The one loop corrections
which contribute ${\cal O}(g^4)$ to the rate fall with $M_H$ (as they
multiply the tree level coupling). However, the one loop
corrections which contribute
${\cal O}(g^6)$ to the rate increase as $M_H$ increases. Hence,
these terms eventually
dominate the rate as $M_H$ becomes large. In Fig.4a the corrected rate
is dominated by the ${\cal O}(g^4)$ terms for small $M_H$,
and hence it
initially falls as $M_H$ increases beyond the kinematic suppression.
Eventually, however, the terms of order ${\cal O}(g^6)$ become
larger than the
${\cal O}(g^4)$ terms and the rate then rises with $M_H$.
This begins to
occur for values of $M_H$ of about 500~GeV.

In Fig.4b we show the rate versus $\tan\beta$ for a Higgs
boson mass of 300~GeV,
a top quark mass of 160~GeV, a squark mass of 1~TeV, and again
for no mixing. We see that the corrected rate is approximately
twice as large as
the tree level value, almost independent of $\tan\beta$.
As we will discuss below, the
rate depends dramatically on $\tan\beta$ once mixing is
included.

In Fig.5 the ratio of the radiatively corrected rate
to the tree level rate is shown versus the top quark mass, for the
same set of
parameters as Fig.4b, and $\tan\beta$=5. Fig.5 illustrates
that the corrected rate depends strongly
on two parameters in the case of no mixing. Clearly the rate
depends on the
value of the top quark mass. But note for $M_H$=1 TeV that
even for a top quark mass as small as
100~GeV the corrected rate is still over a factor of two larger than
at tree level. Thus the relative size of the correction
depends greatly on the value of $M_H$ as well. Note, however,
that when the top quark mass is less than around
120~GeV we expect that the corrections from other sectors will be of
the same order of magnitude as the correction due to the quark/squark
sector included here.

When mixing is included the parameter space increases.  We will
choose a point
in mixing space and examine the effect of mixing in deviations
from that point.
We choose $A-$terms $A_t=A_b=600$~GeV and squark masses
$\tilde{m}_{t_1}=\tilde{m}_{b_1}$=600~GeV,
and $\tilde{m}_{t_2}=\tilde{m}_{b_2}$=300~GeV.
Additionally, we will consider
the two cases $\mu=\pm400$~GeV.
In all three of the figures 6, 7 and 8
the heavy Higgs boson mass is set to 300 GeV and the top
quark mass is 160 GeV.
In order to isolate the effect of mixing we will
plot the ratio of the corrected rate including
mixing to the corrected rate
with no mixing (where the common squark mass is set to 600~GeV).
In Figs.6 we plot this
ratio vs. the squark mass $\tilde{m}_{t_1}$.
We find that the effect due
to mixing is strongly dependent on $\tan\beta$ and $\mu$.
For large values of $\tan\beta$
the effects of mixing are greatly enhanced. As shown in Figs.6, the
inclusion of mixing can change the rate by a factor 1.3
for $\tan\beta$=2 and for
$\tan\beta$=20 by a factor 2.7 or 0.3, for $\mu$=-400~GeV or
$\mu$=+400~GeV, respectively.

Similar ratios are seen in Figs.7,
where the ratio of the corrected rate
including mixing to the corrected rate with no mixing is shown vs.
$A_t$, the top squark mixing parameter. As in Figs.6 the two curves
for $\mu=\pm400$~GeV are similar when $\tan\beta$=2;
the rate can be increased by
50\% or decreased by 25\%. If $\tan\beta$=20 the effects
of mixing are more pronounced and the ratio varies between roughly
1/3 and 3.
The $\mu$=400 GeV curve in Fig.7a (and the $\tan\beta$=2 curve in
Fig.8) does not span the entire
ordinate axis shown because an unphysical region of the squark mixing
parameter space is encountered. In Fig.8 we plot the (mixing) to
(no mixing) ratio vs. the
supersymmetric Higgs mass parameter $\mu$. We see there is little
dependence
on $\mu$ for small $\tan\beta$, while for larger values of
$\tan\beta$ the dependence is
quite significant. If $\tan\beta$=20 the ratio varies
between 4 and 1/36 as $\mu$ varies
from -750 to 750 GeV. Finally, we note that there is very little
dependence
on the bottom squark masses and $A-$term $A_b$ for the mixing
configurations considered.

\section{Conclusions}
To summarize, we have computed the one loop corrections to the decay
rate $\Gamma(H\rightarrow ZZ)$
in the MSSM including third family quark and squark loops.
We perform a Feynman
diagram calculation in the on-mass-shell renormalization scheme.
As the
tree level rate falls like $1/M_H$ for large $M_H$ and we find
corrections that
grow with $M_H$, the corrected rate may be many times the tree
level rate. For
example, at $M_H=1$ TeV the corrected rate may be 13 times the
uncorrected rate for $m_t$=200~GeV (with no squark mixing). The
corrected rate depends very strongly on the squark
mixing parameters. For example, for the mixing configuration
considered here, the rate varies by two orders of magnitude
as the Higgs mass parameter $\mu$ varies between $\pm$750 GeV. Indeed,
the squark mixing parameters $\mu$, $A_t,$ and the top squark masses,
in addition to the top quark mass, must be measured in order to
test the Higgs sector of the MSSM.

\acknowledgements
We would like to thank Mary K. Gaillard for many useful discussions.
One of us (D.P.) acknowledges support from a University of California
at Berkeley Department of Education fellowship.

\unletteredappendix{}
In this Appendix we give explicit analytic expressions
for the self energies,
tadpoles, and form factors introduced in the text.
Our expressions are given in terms of
the standard $A, B, C$ functions introduced by Passarino and Veltman
\cite{Passarino} which appear in one loop calculations. We adopt
the metric (1,-1,-1,-1), which is different than that of
Ref.\cite{Passarino}.
Explicit analytic formulas for these functions appear in
Ref.\cite{'t Hooft}.

To make the equations more concise we adopt the
following conventions. $N_c$ denotes the number of quark colors.
The index $\alpha$ runs over the top and bottom sectors while the
indices $i$, $j$,
and $k$ run over squark mass eigenstates. Thus, $m_\alpha$
denotes a quark mass
while ${\tilde m}_{\alpha i}$ denotes a squark mass.
For the $A$ and $B$ functions we define
$A_{\alpha}=A(m_{\alpha}^2)$,
$\tilde{A}_{\alpha i}=A(\tilde{m}_{\alpha i}^2)$,
$B_{0_{\alpha}}=B_0(p^2,m_{\alpha}^2,m_{\alpha}^2)$,
$\tilde{B}_{0_{\alpha ij}}=B_0(p^2,\tilde{m}_{\alpha i}^2,
\tilde{m}_{\alpha j}^2)$
and similarly for the rest of the $B$'s.
A $C$ function has six arguments:
three external squared momenta and the three squared masses of the
particles which appear in loop of the 3-point diagram.
We thus define
$\tilde{C}_{0_{\alpha ijk}}=C_0(M_Z^2, M_Z^2, M_H^2,
\tilde{m}_{\alpha i}^2,\tilde{m}_{\alpha j}^2,\tilde{m}_{\alpha k}^2)$
and $C_{0_{\alpha}}=C_0(M_Z^2, M_Z^2, M_H^2,
m_{\alpha}^2,m_{\alpha}^2,m_{\alpha}^2)$
with analogous definitions for the rest of the $C$'s.

First we give expressions for the Higgs boson self energies.
\begin{eqnarray}
\Pi_{HH}&&(p^2)=N_c\sum_{\alpha ij}(\tilde{V}_{\alpha ij}^H)^2
\tilde{B}_{0_{\alpha ij}}
-N_c\sum_{\alpha i}\tilde{U}_{\alpha ii}^{HH}\tilde{A}_{\alpha i}
\eqnum{A1}\\
&&-12N_c\sum_\alpha(V_\alpha^H)^2\biggl(m_{\alpha}^2
B_{0_\alpha}+p^2(B_{{21}_\alpha}-B_{1_\alpha})
+{1\over48\pi^2}(m_{\alpha}^2-{p^2\over 6})\biggr).\nonumber
\end{eqnarray}
The various $V$ and $U$ vertex
factors are shown in Fig.9 and explicit expressions appear in
Refs.\cite{Haber Kane,Higgs Hunter}. However, the
$H\;-\; h\;-\; \tilde{q}_{kL}\;-\;\tilde{q}_{kL}$ and
$H\;-\; h\;-\; \tilde{q}_{kR}\;-\; \tilde{q}_{kR}$
vertices given in Ref.\cite{Higgs Hunter} are incorrect. In the
notation of Ref.\cite{Higgs Hunter} the above couplings are
$${ig^2\sin2\alpha\over4}\left(2{T_{3k}-e_k\sin^2
\theta_{\mbox{$\scriptscriptstyle W$}}\over
\cos^2\theta_{\mbox{$\scriptscriptstyle W$}}}
-{m_q^2\over M_W^2}D_k\right)\;\; \mbox{and}\;\;
{ig^2\sin2\alpha\over4}\left(2e_k
\tan^2\theta_{\mbox{$\scriptscriptstyle W$}}
-{m_q^2\over M_W^2}D_k\right)$$
respectively ($D_{up}=1/\sin^2\beta,\;D_{down}=-1/\cos^2\beta$).

$\Pi_{hh}$ is given as $\Pi_{HH}$ with $\tilde{V}_{\alpha ij}^H
\rightarrow\tilde{V}_{\alpha ij}^h,\;\;\tilde{U}_{\alpha ii}^{HH}
\rightarrow\tilde{U}_{\alpha ii}^{hh},$ and $V_\alpha^H
\rightarrow V_\alpha^h$.
$\Pi_{Hh}$ is given as $\Pi_{HH}$ with $(\tilde{V}_{\alpha ij}^H)^2
\rightarrow\tilde{V}_{\alpha ij}^h\tilde{V}_{\alpha ij}^H,\;\;
\tilde{U}_{\alpha ii}^{HH}\rightarrow\tilde{U}_{\alpha ii}^{Hh},$ and
$(V_\alpha^H)^2 \rightarrow V_\alpha^H V_\alpha^h.
\;\;\Pi_{AA}$ is given as $\Pi_{HH}$ with
$\tilde{V}_{\alpha ij}^H\rightarrow\tilde{V}_{\alpha ij}^A,\;\;
\tilde{U}_{\alpha ii}^{HH}\rightarrow\tilde{U}_{\alpha ii}^{AA},
\;\;V_\alpha^H \rightarrow V_\alpha^A,$ and $B_{0_\alpha}
\rightarrow{1\over3}B_{0_\alpha}.$
Next we list the transverse part of
the gauge boson self energies.
\begin{eqnarray}
&&\Pi_{ZZ}^T(p^2)\;=\;N_c\sum_{\alpha i}
\tilde{U}_{\alpha ii}^{ZZ} \tilde{A}_{\alpha i}\eqnum{A2}\\
&&\;-\;2N_c\sum_{\alpha ij} (\tilde{V}_{\alpha ij}^Z)^2
\biggl(\tilde{m}_{\alpha i}^2\tilde{B}_{0_{\alpha ij}} -
(\tilde{m}_{\alpha i}^2-\tilde{m}_{\alpha j}^2
+p^2)\tilde{B}_{1_{\alpha ij}}
+ p^2\tilde{B}_{{21}_{\alpha ij}}
+{1\over16\pi^2}({\tilde{m}_{\alpha i}^2
+\tilde{m}_{\alpha j}^2\over2}-
{p^2\over6})\biggr)\nonumber \\
&&\;-\;8N_c\sum_\alpha
\biggl( (V_{5\alpha}^Z)^2m_{\alpha}^2 B_{0_{\alpha}}
+ \left((V_\alpha^Z)^2+(V_{5\alpha}^Z)^2\right)p^2(B_{{21}_\alpha}-
B_{1_\alpha}) \biggr) \nonumber
\end{eqnarray}
\begin{eqnarray}
&&\Pi_{WW}^T(p^2)\;=\;N_c\sum_{\alpha i}\tilde{U}_{\alpha ii}^{WW}
\tilde{A}_{\alpha i}\eqnum{A3}\\
&& \;-\;2N_c\sum_{\alpha ij}
    (\tilde{V}_{ij}^W)^2\biggl(\tilde{m}^2_{ti}\tilde{B}^W_0
    -(\tilde{m}^2_{ti}-\tilde{m}^2_{bi}+p^2)\tilde{B}^W_1
+p^2\tilde{B}^W_{21}
 +{1\over 16\pi^2}({\tilde{m}^2_{ti}+\tilde{m}^2_{bi}\over2}-
  {p^2\over6})\biggr)\nonumber\\
&&\;-\;8N_c(V^W)^2\biggl( m_t^2B^W_0-(m_{t}^2-m_{b}^2+2p^2)B^W_1
     +2p^2B^W_{21}\biggr) \nonumber
\end{eqnarray}
where $\tilde{B}^W = B(p^2,{\tilde m}_t^2,{\tilde m}_b^2)$ and
$B^W = B(p^2,m_t^2,m_b^2)$.
$\Pi^T_{\gamma\gamma}$ is given as $\Pi_{ZZ}^T$ with
$\tilde{V}_{\alpha ij}^Z\rightarrow\tilde{V}_{\alpha ij}^\gamma,\;
\tilde{U}_{\alpha ii}^{ZZ}\rightarrow
\tilde{U}_{\alpha ii}^{\gamma\gamma},\;
V_\alpha^Z\rightarrow V_\alpha^\gamma,\;\;V_{5\alpha}^Z\rightarrow 0$.
The heavy Higgs boson tadpole contribution is given by
\begin{equation}
T_H=N_c\biggl(4\sum_\alpha V_\alpha^H
m_{\alpha}A_{\alpha}-\sum_{\alpha i}\tilde{V}_{\alpha ii}^H
\tilde{A}_{\alpha i}\biggr).
\eqnum{A4}
\end{equation}
$T_h$ is given as $T_H$ with $\tilde{V}_{\alpha ii}^H\rightarrow
\tilde{V}_{\alpha ii}^h$ and $V_\alpha^H \rightarrow V_\alpha^h$.
Lastly, the two three-point Feynman diagram form factors which are
relevant for
calculating the Higgs boson decay rate are
\pagebreak
\begin{eqnarray}
D_0=8N_c&&\sum_\alpha m_\alpha V_\alpha^H((V_\alpha^Z)^2
+(V_{5\alpha}^Z)^2)\nonumber\\
&&\times\biggl(4C_{{24}_\alpha}
+(M_H^2-2M_Z^2)C_{{12}_\alpha}
+ 2M_Z^2C_{{11}_\alpha} -{M_H^2\over2}C_{0_\alpha}
- B_0(M_Z^2,m_{\alpha}^2,
m_{\alpha}^2)\biggr)\nonumber \\
&&+\;\;8N_c\sum_\alpha m_{\alpha} V_\alpha^H
(V_{5\alpha}^Z)^2\biggl(\left(M_H^2-4m_{\alpha}^2\right)C_{0_\alpha}
-2B_0(M_Z^2,m_{\alpha}^2,m_{\alpha}^2)\biggr)\eqnum{A5}\\
&&-\;\;8N_c\sum_{\alpha ijk}\tilde{V}_{\alpha ki}^H
\tilde{V}_{\alpha ij}^Z\tilde{V}_{\alpha jk}^Z
\tilde{C}_{{24}_{\alpha ijk}}
\;\;+\;\;
N_c\sum_{\alpha ij}\tilde{V}_{\alpha ij}^H\tilde{U}_{\alpha ij}^{ZZ}
B_0(M_H^2,\tilde{m}_{\alpha i}^2,\tilde{m}_{\alpha j}^2)\nonumber
\end{eqnarray}
and
\begin{eqnarray}
D_4=&&8N_c\sum_\alpha m_\alpha V_\alpha^H
\biggl(\left((V_\alpha^Z)^2+(V_{5\alpha}^Z)^2\right)
\left(4C_{{23}_\alpha}+C_{0_\alpha}-4C_{{12}_\alpha}\right)
-2(V_{5\alpha}^Z)^2\left(C_{{11}_\alpha}-C_{{12}_\alpha}\right)
\biggr)\nonumber \\
&&\;-\;8N_c\sum_{\alpha ijk}\tilde{V}_{\alpha ki}^H
\tilde{V}_{\alpha ij}^Z\tilde{V}_{\alpha jk}^Z
\left(\tilde{C}_{{23}_{\alpha ijk}}
-\tilde{C}_{{12}_{\alpha ijk}}\right).
\eqnum{A6}
\end{eqnarray}

\newpage

\figure{The Higgs boson tadpole diagram.}
\figure{The Higgs boson and gauge boson self energy diagrams.}
\figure{The one loop $H\rightarrow ZZ$ Feynman diagrams included
in our calculation. The dashed loops represent squarks, the solid loop
represents quarks.}
\figure{The tree level and one loop corrected
decay rate $\Gamma(H\rightarrow ZZ)$. Fig.4a shows the
rates vs. the heavy Higgs boson mass for $\tan\beta$~=5,
$m_t$=160~GeV, and no mixing. Fig.4b shows the decay rates vs.
$\tan\beta$
for the same parameters as in Fig.4a, with $M_H$=300~GeV.}
\figure{The ratio of the corrected decay rate to the tree level rate
vs. the top quark mass. The parameters are the same as in Fig.4a,
with the Higgs boson mass set to 300~GeV and 1~TeV.}
\figure{The ratio of the corrected rate including mixing
to the corrected rate without mixing vs. the top squark mass
$\tilde{m}_{t_1}$ as explained in the text. The top quark
mass is 160~GeV.}
\figure{The ratio of the corrected rate including mixing
to the corrected rate without mixing vs. the top mixing parameter
$A_t$, as explained in the text. The
top quark mass is 160~GeV. In Fig.7a the point where the
$\mu$=400~GeV curve stops
corresponds to an unphysical point in squark mixing parameter space.}
\figure{The ratio of the corrected rate including mixing
to the corrected rate without mixing, as explained in the text, vs.
$\mu$. The heavy Higgs boson mass is 300~GeV, and the top quark
mass is 160~GeV. The curve for $\tan\beta=2$ stops
at an unphysical point in the squark mixing parameter space.}
\figure{The vertices used in the Appendix are displayed. The values
of the vertex factors may be found in
Refs.\cite{Haber Kane,Higgs Hunter}. Note
that the value of $U^{Hh}_{\alpha ij}$ listed in
Ref.\cite{Higgs Hunter}
is incorrect (see the Appendix for the correct values of these
vertices).}

\end{document}